\documentclass[twocolumn,showpacs,prl]{revtex4}
\usepackage{graphicx}
\usepackage{dcolumn}
\usepackage{bm}
\newcommand{\oncite}{\onlinecite}

\begin{document}

\title {Bose-Fermi solid and its quantum melting in an one-dimensional optical lattice}
\author{Bin Wang$^1$}
\author{Daw-Wei Wang$^2$}
\author{S. Das Sarma$^1$}
\affiliation{$^1$Condensed Matter Theory Center, Department of
Physics, University of Maryland, College Park, Maryland 20742,
USA
\\
$^2$Physics Department and NCTS, National Tsing-Hua
University, Hsinchu 30013, Taiwan}

\begin{abstract}
We investigate the quantum phase diagram of Bose-Fermi mixtures of
ultracold dipolar particles trapped in one-dimensional optical
lattices in the thermodynamic limit. With the presence of
nearest-neighbor (N.N.) interactions, a long-ranged ordered
crystalline phase (Bose-Fermi solid) is found stabilized between a
Mott insulator of bosons and a band-insulator of fermions in the
limit of weak inter-site tunneling ($J$). When $J$ is increased,
such a Bose-Fermi solid can be quantum melted into a Bose-Fermi
liquid through either a two-stage or a three-stage transition,
depending on whether the crystalline order is dominated by the
N.N. interaction between fermions or bosons. These properties can
be understood as quantum competition between a pseudo-spin
frustration and a pseudo-spin-charge separation, qualitatively
different from the classical picture of solid-liquid phase
transition.
\end{abstract}

 \maketitle

The successful experimental preparation of ultracold dipolar
atoms\cite{dipolarexp} and polar molecules\cite{polarmolexp} has
opened a new direction of strongly correlated physics. The
long-ranged anisotropic feature of the dipolar interaction can
lead to qualitatively different physics as compared to the
conventional ultracold atom systems. Recent theoretical efforts
have been mostly devoted to either bosonic or fermionic dipolar
gases \cite{fermipolar,dipolarlattice,Altman}. However, not much
work is reported on dipolar Bose-Fermi mixtures, although their
counterparts in conventional ultracold atoms have been
extensively studied both experimentally \cite{BFexp} and
theoretically \cite{BFtheory,Mathey04,Pollet06,Sengupta07}. When
the dipolar interaction is included, some crystalline order may be
realized [\oncite{dipolarlattice}], providing completely new
insight onto the solid-liquid phase transition, which is one of
the most important subjects in classical physics. We
show in this work that a Bose-Fermi dipolar mixture allows
for extremely rich interaction physics by manifesting a complex
and experimentally accessible quantum phase diagram containing
many exotic quantum phases hitherto not considered in cold atomic
gases.

In this paper, we study various kinds of Bose-Fermi mixtures of
dipolar particles loaded in an one dimensional (1D) optical
lattice. Keeping only the on-site and the nearest-neighbor (N.N.)
interactions, five distinct quantum phases can be identified:
Bose-Fermi solid (BFS), density wave Bose-Fermi Mott state
(DW-BFM), density wave Bose-Fermi liquid (DW-BFL), uniform
Bose-Fermi Mott state (UBFM), and uniform Bose-Fermi liquid
(UBFL), as shown in Fig. 1. The BFS phase has alternating boson
and fermion densities similar to its classical counterpart, but
can be quantum melted even at $J=0$ through an Ising-type
frustration when the dipole moments of bosons and fermions are
equal. When $J$ increases, the BFS state can have another quantum
melting through a multi-stage transition, which can be understood
as Kosterlist-Thouless type with the pseudo-spin-charge
separation[\oncite{KT}]. We also discuss the experimental
preparation and observation of these quantum phases, and connect
our results to the metallic electronic reconstruction at the
interface of a band insulator and a Mott insulator
[\oncite{Millis}].

\begin{figure}
\includegraphics [width=8.5 cm] {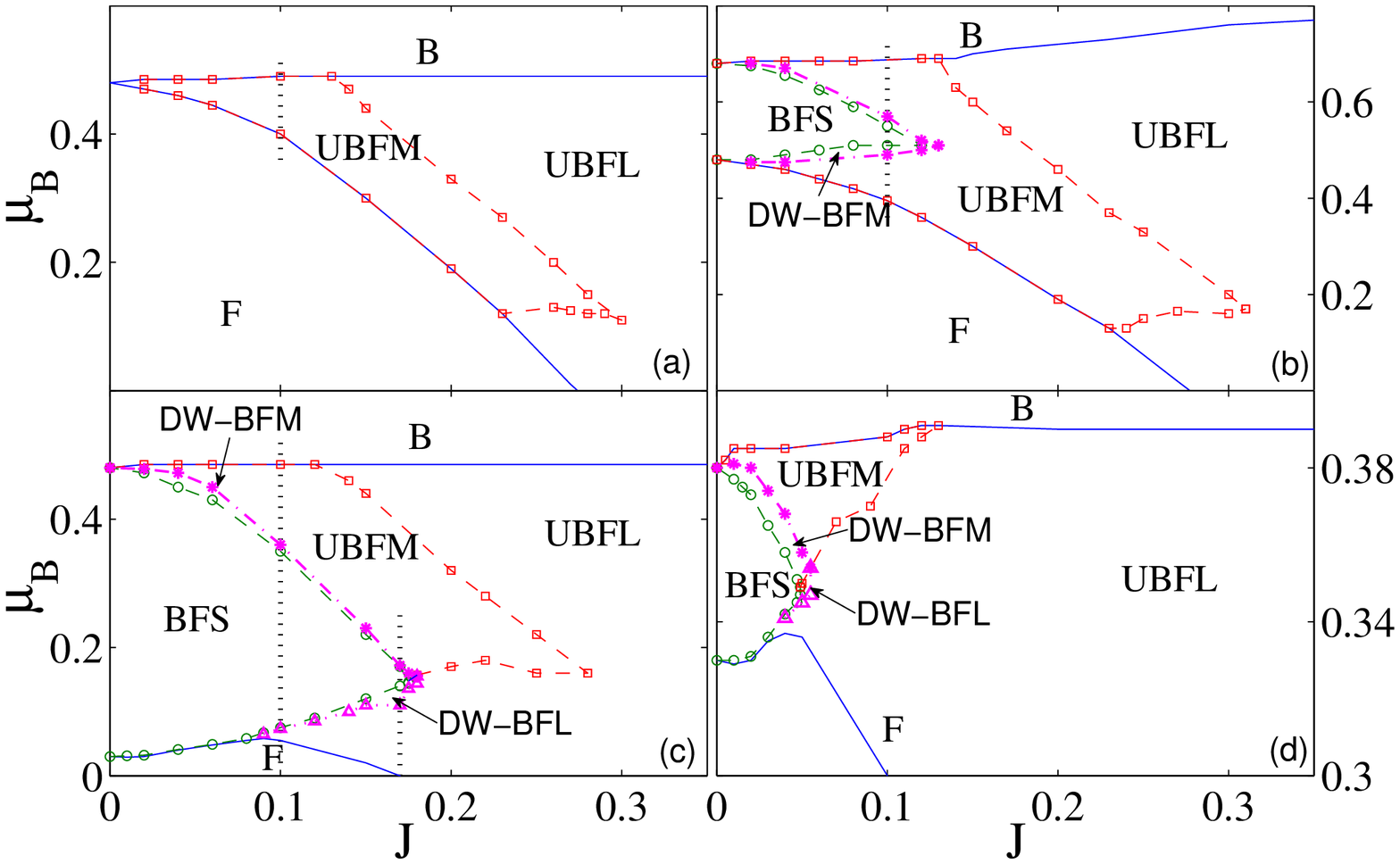}
\caption[{[Fig.1]}]{Phase diagrams of dipolar Bose-Fermi mixtures:
(a)-(d) corresponds to Case I-IV as defined in the text. $V_{BB}=0.1$ in (b) and (d),
and $V_{FF}=0.225$ in (c) and (d). $V_{BF}=0.15$ in (d) and
$\mu_F=0.48$ for all panels. $U=1$ is the energy unit. The vertical dotted lines in (a)-(c) show
the intersect for which the particle density is presented in Fig. 3.}
\end{figure}

To describe our model system, we adopt the following extended
Bose-Fermi Hubbard Hamiltonian: $H=H_B+H_F+H_{BF}$, where
\begin{eqnarray}
\nonumber H_B&=&-J_{B}\sum_{\langle
i,j\rangle}b^{\dagger}_{i}b_{j}+\frac{U_{BB}}{2}\sum_i
n^B_{i}(n^B_{i}-1)\\
\nonumber
& &+\frac{V_{BB}}{2}\sum_{\langle i,j\rangle}n^B_{i}n^B_{j}-\mu_B\sum_i
n^B_{i},\\
\nonumber H_F&=&-J_{F}\sum_{\langle
i,j\rangle}f^{\dagger}_{i}f_{j}+\frac{V_{FF}}{2}\sum_{\langle
i,j\rangle}n^F_{i}n^F_{j}-\mu_F\sum_i
n^F_{i},\\
H_{BF}&=&U_{BF}\sum_{i}n^B_{i}n^F_{i}+V_{BF}\sum_{\langle
i,j\rangle}n^B_{i}n^F_{j}.
\label{H}
\end{eqnarray}
Here, $b_i$ and $f_i$ are the bosonic and fermionic field
operators at the $i$th site. $n^B_i=b^{\dagger}_ib_i$ and
$n^F_i=f^{\dagger}_if_i$ are density operators. $J_{a}$ and
$\mu_{a}$ are respectively the hopping rates between adjacent
sites ($\langle i,j\rangle$) and chemical potentials; $U_{ab}$ and
$V_{ab}$ describe the on-site and the N.N. interactions
respectively, with $a$ and $b$ denoting bosons ($B$) or fermions
($F$). Such a system can in principle be engineered in ultracold
dipolar gases. Longer range interactions are neglected here
because they do not qualitatively change the phase diagram near
half filling \cite{2Dpolarphase}.

In the hard-core boson limit ($U_{BB}\to\infty$) one can map the
bosonic operators to fermionic ones, and the whole system can be
transformed to a special extended Hubbard model with
spin-dependent N.N. interactions. Such a system breaks the SU(2)
symmetry and cannot be solved analytically. Since we are
interested here in the interplay of quantum statistics with N.N.
interactions, we will consider a soft-core case by setting $J_a=J$
and $U_{ab}\equiv U=1$ for simplicity. Near the Mott insulator
regime, we can reduce the Hilbert space into 0, 1 and 2 particles
per site and hence such soft-core bosons can also be mapped to a
spin-1 chain \cite{Altman}, i.e. Eq. (\ref{H}) can be mapped onto
a unique spin ladder composed of a spin-one chain and a spin-half
chain with frustrated anti-ferromagnetic coupling as $V_{ab}>0$.
In our calculation, we stick with the particle model and the
ground states are obtained using an infinite lattice version of
the time-evolving-block-decimation (TEBD) algorithm \cite{TEBD},
which exploits a tensor product representation of many-body states
and is especially efficient for the gapped solid phase that we are
interested in here. 

In order to clarify the effects of the N.N. interactions, we
calculate and compare the following six representative cases: Case
I: neither bosons or fermions are dipolar particles, i.e.
$V_{ab}=0$; Case II: only bosons are dipolar particles, i.e.
$V_{BB}>0$ and $V_{BF}=V_{FF}=0$; Case III: only fermions are
dipolar particles, i.e. $V_{FF}>0$ and $V_{BB}=V_{BF}=0$; Case IV:
both bosons and fermions are dipolar particles, i.e.
$V_{BB}/V_{BF}=V_{BF}/V_{FF}\equiv\chi$ and $\chi=2/3<1$; Case V:
same as Case IV but for $\chi=3/2>1$; Case VI: same as Case IV but
for $\chi=1$. Several representative quantum phase diagrams are
shown in Fig. 1, where panels (a)-(d) correspond to Case I-IV
respectively. Results of Case V and Case VI are qualitatively the
same as Case II and Case I respectively. To save space, we do not
show their phase diagrams here.
\begin{figure}
\includegraphics [width=8 cm] {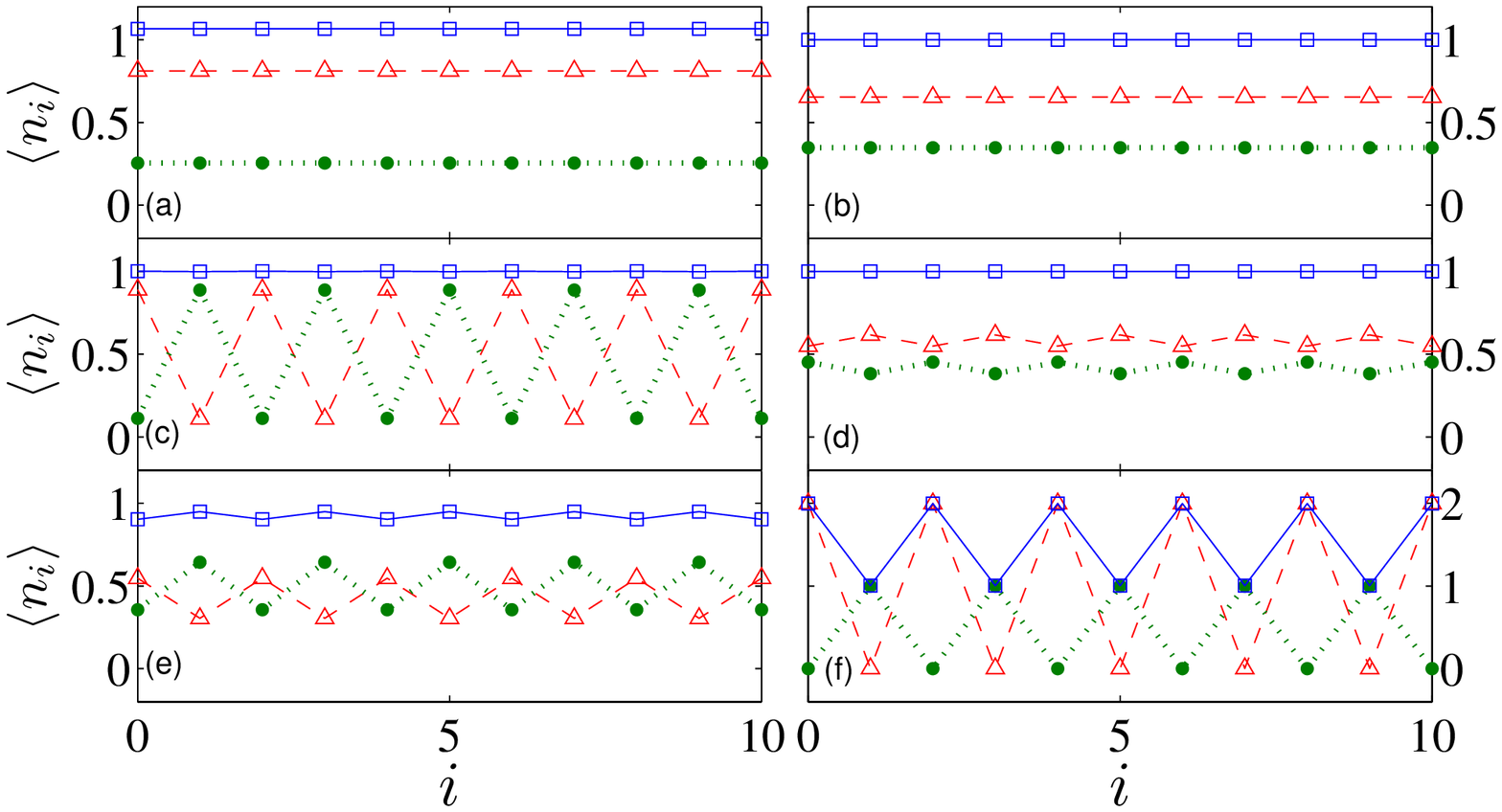}
\caption[{[Fig.2]}]{Typical real space distributions of boson
(open triangle), fermion (filled circle), and total density (open
diamond) in (a) UBFL, (b) UBFM, (c) BFS, (d) DW-BFM, and (e)
DW-BFL phases. Panel (f) shows a BFS phase with higher filling
fraction of bosons. }
\end{figure}

In the limit of small(large) $\mu_B$, the system is a pure
Fermi(Bose) gas, denoted by $F(B)$. Within these parameters, the
ground state can be a bosonic Mott insulator or a 1D Bose liquid
in regime B, while in regime F it can be a band insulator or a
Fermi liquid. In the mixed regime, there are five distinct quantum
phases to be observed: (i) in the large $J$ regime, we have a UBFL
phase, whose quasi-long-ranged order can be obtained by using
bosonization[\oncite{Mathey04}]. The density distribution is
uniform in space for each species, as shown in Fig. 2(a). (ii) In
the regime of stronger interaction, there is a UBFM phase, where
the total filling fraction is found to be unity through the entire
regime with a gapped excitation. The UBFM phase exists even
without N.N. interaction (Fig. 1(a)) \cite{Sengupta07}, and it has
a uniform and compressible inter-species density difference (Fig.
2(b)). (iii) When the N.N. tunneling is close to zero, we find a
BFS phase, where both species are at half-filling but have an
alternating density pattern (see Fig. 2(c)). Such a phase is
incompressible for both species (i.e. $\partial \langle
n^a_i\rangle/\partial\mu_{b}=0$ for $a,b=\{B,F\}$), while their
density oscillation amplitudes decrease as $J$ increases. (iv)
Near the boundary between BFS and UBFM, there is a narrow regime,
where the BFM phase has a density wave order (DW-BFM), i.e.
$\langle n_i^B-n_i^F\rangle$ oscillates periodically while the
total filling remains integer (see Fig. 2(d)). (v) Finally, near
the boundary between BFS and BFL, we find a BFL with a density
wave order in both species (named DW-BFL). Different from DW-BFM,
DW-BFL is compressible with an incommensurate total filling
fraction (Fig. 2(e)). This phase can also be regarded as a
supersolid phase in the quasi-long-ranged order sense.

In Fig. 2(a)-(e), we show typical spatial density distributions
for BFL, UBFM, BFS, DW-BFM, and DW-BFL phases respectively. We
note that, although the true spin-charge separation exists only in
the low energy regime \cite{Mathey04}, we can still
phenomenologically define the pseudo-charge to be $\langle
n^B_i+n^F_i\rangle$, and pseudo-spin to be $\langle
n^B_i-n^F_i\rangle$. As a result, the BFL phase has gapless
excitations in both pseudo-charge and pseudo-spin sectors, UBFM
has a gapped charge excitation, and BFS has gapped excitation in
both sectors. DW-BFM has a gapped pseudo-charge excitation with a
gapless pseudo-spin-density wave order, while DW-BFL has two
gapless excitations on top of the pseudo-spin density wave order.
Note that our numerical calculation is
for an infinite 1D system, so the observed density wave order is
always a truly long-ranged order. In Fig. 3(a)-(c), we further
plot the average filling fraction of bosons and fermions (together
with the total fraction) as a function of the bosonic chemical
potential $\mu_B$, showing the regime of different phases. The
plateau for particle density at 0.5 in panels (b) and (c)
corresponds to the half-filled BFS phase. The plateau of total
density at unity can be identified to be a Mott phase if densities
of individual species vary as a function of the chemical
potential. Finally, the regime with a varying total density
corresponds to the BFL phase. UBFM and DW-BFM cannot be
distinguished here (same as UBFL and DW-BFL) since only the
average filling fraction is shown. In Fig. 3(d), the density wave
order is plotted as a function of $\mu_B$ at $J=0.17U$, and is
found to disappear continuously when changing $\mu_B$
away from BFS regime.

Note that the origin of our BFS state is different from that of
the Neel(Ising) state predicted in \cite{Pollet06}, where no N.N.
interaction exists. In the latter case, the effective long-ranged
interaction between heavier species is induced by the quantum
fluctuations of the lighter species, and becomes effectively
vanishing when $J_B=J_F$. Besides, we also note that at higher
$\mu_B$ (outside the regime of Fig. 1), a BFS state with more than
one particle per site is also possible (see Fig. 2(f)), but we
will not discuss such phases here due to space limitation.
\begin{figure}
\includegraphics [width=8.5 cm] {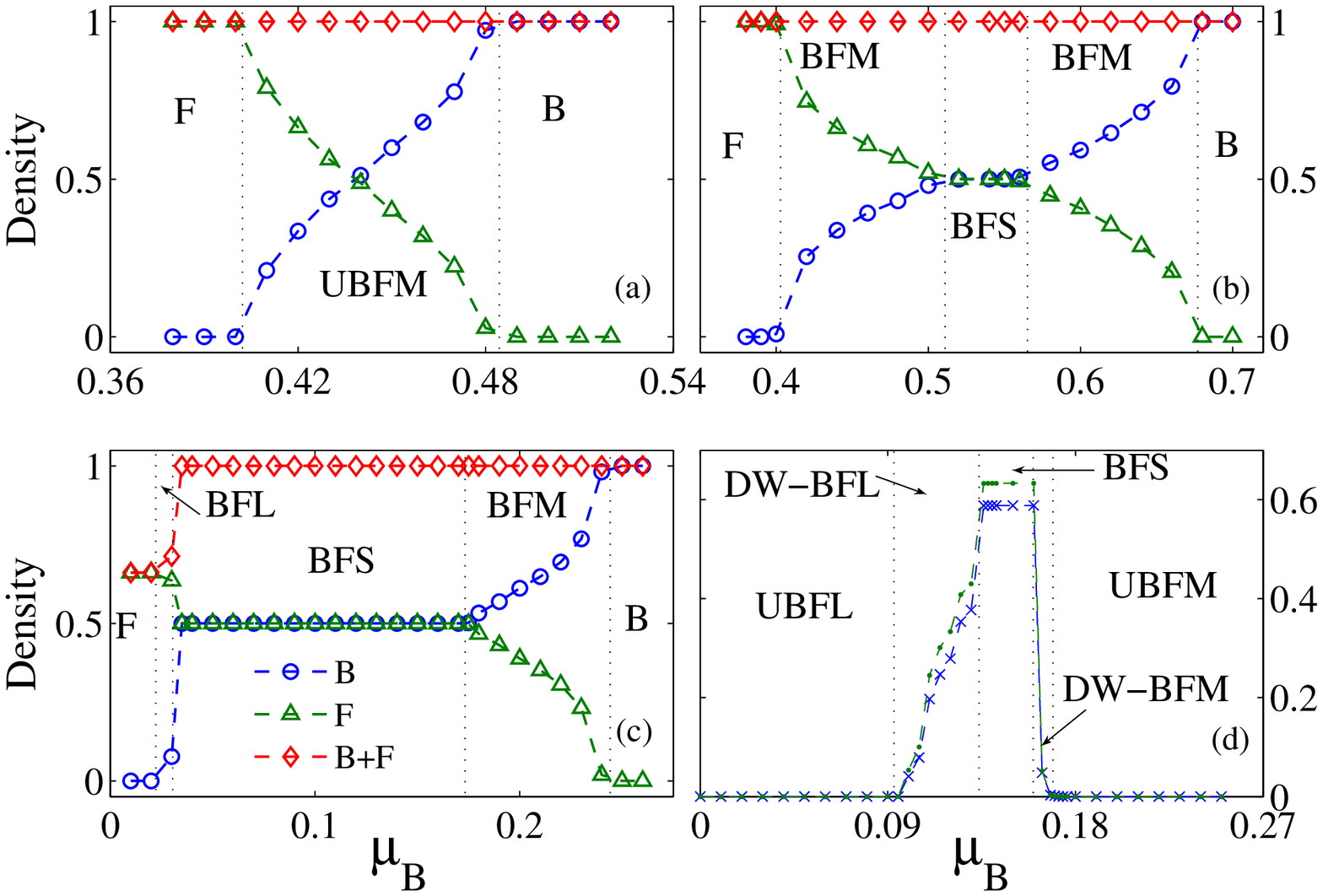}
\caption[{[Fig.3}]{(a)-(c) Average boson (circle), fermion
(triangle), and total (diamond) densities as functions of $\mu_B$
at $J=0.1$. All other parameters are the same as in Fig. 1(a)-(c)
respectively. (d) shows the density wave order at $J=0.17$
for the same system as in (c). The solid
dot and check markers correspond to the density wave order,
$|\langle n^B_i\rangle-\langle n^B_{i+1}\rangle|$ and $|\langle n^F_i\rangle-\langle
n^F_{i+1}\rangle|$.}
\end{figure}

To gain more insight on the crystalline phase, it is instructive
to look into the classical limit at $J=0$, where no difference
between fermionic and bosonic particles is expected. By comparing
the energies of various configurations, for example, half-filled
Bose/Fermi density wave, Bose Mott Insulator, and half-filled BFS,
we can obtain the necessary condition for a BFS phase to be:
$\mu_B,\mu_F>2V_{BF}$ and
\begin{eqnarray}
2(V_{FF}-V_{BF})>\mu_F-\mu_B>2(V_{BF}-V_{BB}). \label{condition}
\end{eqnarray}
It is therefore interesting to find that for a symmetric case,
i.e. Case VI and $V_{ab}=V$, a crystal exists only when
$\mu_F=\mu_B>2V$, and the state $\{\cdots BFBF\cdots\}$ would be
degenerate with all states of one particle per site. This can also
be understood in the spin-ladder picture we described earlier: the
N.N. interactions, $V_{ab}>0$, provide an Ising-type frustration,
namely, $V_{BF}$ competes with $V_{BB}$ and $V_{FF}$, and hence
smear out the anti-ferromagnetic order. At a finite but small $J$,
the degeneracy is removed, but the density wave order is still so
small that the BFS phase just disappears when $\mu_B$ slightly
differs from $\mu_F$. Therefore, according to our numerical
calculation and conditions in Eq. (2), we conclude that the BFS
phase is unlikely to be observed in a realistic experimental
situation for the symmetric case ($\chi=1$) due to the frustrated
nature of the dipolar mixture. The system is therefore still
compressible and uniform in density, as the UBFM case shown in
Fig. 1(a).

Away from the symmetric case, we first consider the pro-boson
case, i.e. the N.N. interaction between bosons is stronger than
that between fermions ($\chi>1$) and hence dominates the crystal
order formation. For such a situation, we show our results for
Case II in Fig. 1(b). Results for Case V are similar. Such
pro-boson BFS is quantum melted to the UBFL via a three-stage
process: first, the ground state has a second order transition
toward the DW-BFM phase, having a compressible pseudo-spin density
wave order on top of a pseudo-charge Mott phase. When $J$ becomes
slightly larger, the density wave order disappears and the ground
state becomes a UBFM phase, where the single particle Green's
function decays exponentially (see Fig. 4). Finally, the
pseudo-charge gap of UBFM becomes zero at larger $J$, leading to a
BFL phase. The last two transitions should belong to the
Kosterlitz-Thouless universality class in the pseudo-spin and
pseudo-charge sectors respectively, as described by an X-Y model
in the 1+1 dimensional system.
\begin{figure}
\includegraphics [width=8.5 cm] {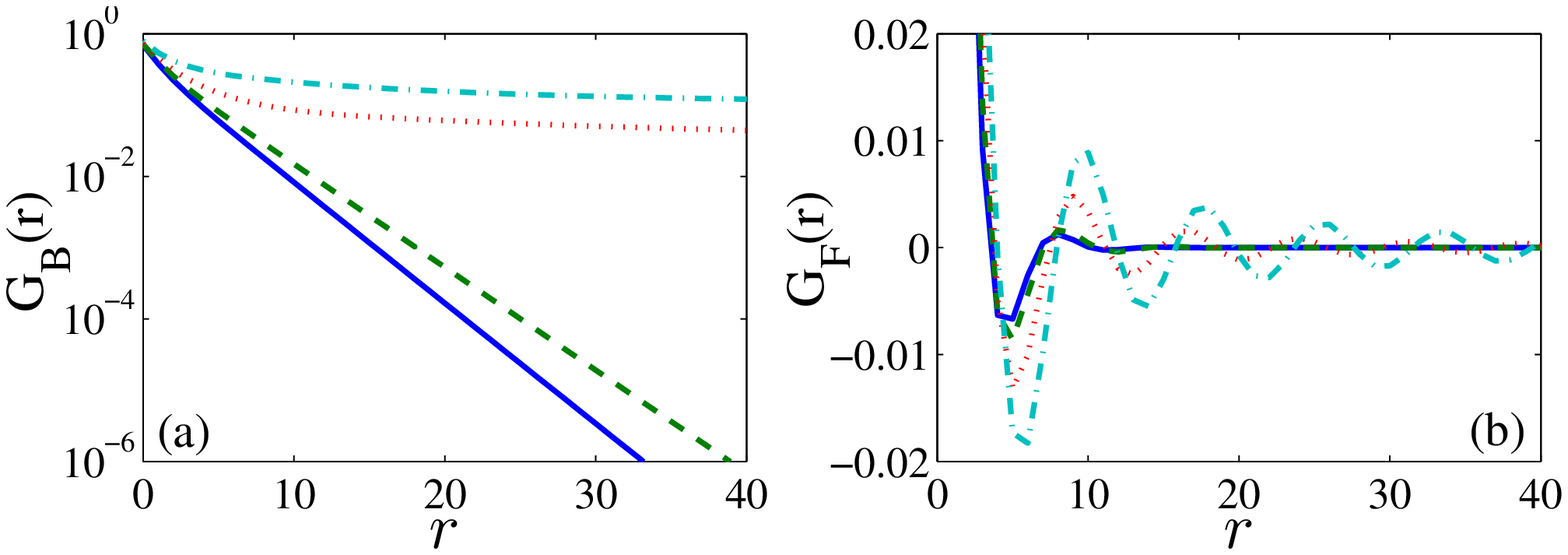}
\caption[{[Fig.4]}]{Green functions for (a) bosons and (b) fermions
in a few samples inside the UBFM and UBFL phase. $J=0.15$, $0.16$, $0.17$, and $0.18$ for
the solid, dashed, dotted, and dash-dotted lines respectively. $\mu_B=0.4$ and
all other parameters are the same as in Fig. 1(c).}
\end{figure}

However, such kind of three-stage process from BFS to BFL does not
always exist for a pro-fermion case, i.e. when the crystal order
is mainly provided by the N.N. interaction between fermions. We
provide the quantum phase diagrams for both Case III and Case IV
in Figs. 1(c) and (d). In Fig. 1(c), we find that in addition to
the three-stage melting process discussed above, in the lower
portion of the BFS boundary, there is a different melting process:
first from BFS to DW-BFL and then from DW-BFL to UBFL.

Such different melting processes for pro-boson and pro-fermion
solids demonstrate the quantum character of BFS. Their common
feature is the smooth decrease of the density wave order when
outside the BFS phase, while the difference lies in that the
pseudo-charge gap disappears together with the pseudo-spin gap in
the lower portion of BFS boundary for the pro-fermion case. This
result can be understood when comparing the BFS regime in Fig.
1(b) and (c). First, without N.N. interaction (Fig. 1(a)), the BFS
is degenerate with other states at a single point, $\mu_B=\mu_F$
and $J=0$. When $V_{BB}>0$ and other $V_{ab}=0$, from the
condition shown in Eq. (\ref{condition}), the regime of BFS is
$0<\mu_B-\mu_F<2V_{BB}$, i.e. the BFS regime is entirely inside
the original Mott insulator regime. As a result, when BFS is
melted, the whole system is still in the regime of commensurate
total filling fraction, and a Mott phase with a pseudo-charge gap
is then expected in the strong interaction limit. On the other
hand, when only $V_{FF}>0$ and other $V_{ab}=0$, the condition for
BFS at $J=0$ gives $0<\mu_F-\mu_B<2V_{FF}$, which is inside the
regime of fermion band insulator as shown in Fig. 1(a). Unlike
Mott insulator for bosonic particles, the band insulator of
fermions exists in non-interacting limit and hence can be easily
destroyed by finite interactions, leading to a Luttinger liquid in
1D with incommensurate filling. That is why both pseudo-spin and
pseudo-charge gaps disappear in the DW-BFL and UBFL regimes when
BFS is melted in the lower boundary of BFS phase in Fig. 1(c). The
nature of such a transition is still unclear. Finally, from Fig.
1(d), we can see that the phase diagram for such a pro-fermion
case is qualitatively similar to Fig. 1(c), but the finite value
of $V_{BF}$ can suppress the BFS phase as well as the BFM phase,
consistent with the fact that no BFS exists in the symmetric case
due to Ising-type frustration.

In Fig. 4 we also present the single particle Green's function of
(a) bosons: $G_B(r)=\langle b_i^{\dagger}b_{i+r}\rangle$ and (b)
the fermions, $G_F(r)=\langle f_i^{\dagger}f_{i+r}\rangle$. It
clearly shows that the single particle correlation of bosons and
fermions has an exponential decay in the UBFM regime (solid and
dashed lines), while they have a slower algebraic decay due to the
quasi-long-ranged order inside the UBFL regime (dotted and
dashed-dotted lines).

Before concluding, we note that the system we discuss here can be
easily realized by a mixture of dipolar atoms/molecules loaded
into a 1D optical lattice. One may provide two separate trapping
potentials for bosons and fermions, and prepare them to be a Mott
insulator and a band insulator respectively. When the relative
distance between these two traps is getting smaller, the
inter-mediate regime between these two clouds experiences a
continuous change of chemical potentials, making a potential
realization of BFS and other exotic phases discussed here. Our
results indicate that observing a BFS could be very challenging
for systems in which bosons and fermions have close dipole moments
(say ${}^{39}$K${}^{87}$Rb-${}^{40}$K${}^{87}$Rb or
$^{52}$Cr-$^{53}$Cr), due to strong frustration. It is therefore
more promising to consider a mixture of dipolar and non-dipolar
particles to realize a BFS. It is interesting to note that,
compared to the metallic and ferromagnetic interface of an
electronic band insulator and a Mott insulator [\oncite{Millis}],
the intermediate BFS phase here is an insulator with
anti-ferromagnetic order, which can be easily observed in a
time-of-flight measurement.

In summary, we have studied the ground-state phase diagram of the
Bose-Fermi Hubbard model with arbitrary N.N. interactions. We
discover the possibility and the condition to find a novel
Bose-Fermi solid phase, which has a nontrivial quantum melting due
to the mixture of quantum statistics even in 1D. This system can
be experimentally realized within the present technique.

We acknowledge fruitful discussion with H.-H. Lin. DWW appreciates
the hospitality of JQI during the initial discussion of this work.
This work is supported by JQI AFOSR MURI.



\begin{thebibliography} {99}
\bibitem{KT}
The 2D classical melting is described by the seminal Kosterlitz-Thouless-Halperin-Nelson-Young (KTHNY) theory, which
suggests a two-stage process mediated by the pairwise creation of topological defects.
\end{thebibliography}
\end{document}